# Angle-dependence and optimal design for magnetic bubblecade with maximum speed

Duck-Ho Kim,[1,†] Kyoung-Woong Moon,[2] Sang-Cheol Yoo,[1,3] Dae-Yun Kim,[1] Byoung-Chul Min,[3] Chanyong Hwang,[2] and Sug-Bong Choe[1,★]

[1]Department of Physics and Institute of Applied Physics, Seoul National University, Seoul, 08826, Republic of Korea.

[2]Center for Nanometrology, Korea Research Institute of Standards and Science, Daejeon, 34113, Republic of Korea.

[3]Center for Spintronics, Korea Institute of Science and Technology, Seoul, 02792, Republic of Korea.

[†]Present Address: Institute for Chemical Research, Kyoto University, Kyoto, Japan

[★]Correspondence to: sugbong@snu.ac.kr

**Unidirectional magnetic domain-wall motion is a key concept underlying next-generation application devices. Such motion has been recently demonstrated by applying an alternating magnetic field, resulting in the coherent unidirectional motion of magnetic bubbles. Here we report the optimal configuration of applied magnetic field for the magnetic bubblecade—the coherent unidirectional motion of magnetic bubbles—driven by a tilted alternating magnetic field. The tilted alternating magnetic field induces asymmetric expansion and shrinkage of the magnetic bubbles under the influence of the Dzyaloshinskii-Moriya interaction, resulting in continuous shift of the bubbles in time.**

**By realizing the magnetic bubblecade in Pt/Co/Pt films, we find that the bubblecade speed is sensitive to the tilt angle with a maximum at an angle, which can be explained well by a simple analytical form within the context of the domain-wall creep theory. A simplified analytic formula for the angle for maximum speed is then given as a function of the amplitude of the alternating magnetic field. The present observation provides a practical design rule for memory and logic devices based on the magnetic bubblecade.**

Magnetic domain-wall (DW) motion has been intensively studied as a test body of the emerging spin-dependent phenomena [1–5] as well as a building block of the potential application devices [6–9]. Such DW motion has been achieved by the spin-orbit [2–4,10] or spin-transfer [11–14] torques through injection of the spin-polarized current. Fairly recently, Moon *et al*. [9] proposed another scheme to generate a similar motion by applying an alternating magnetic field to chiral DWs. The coherent unidirectional bubble motions generated by this scheme is referred as a "magnetic bubblecade," which enables the demonstration of multi-bit bubble memory operation. The key concept underlying this scheme relies on the broken symmetry caused by the Dzyaloshinskii–Moriya interaction (DMI), which induces the asymmetric expansion and shrinkage of magnetic bubbles [9,15,16]. This scheme was further revealed to induce the coherent unidirectional motion of the DWs and skyrmions [17]. Here, we investigate the optimal angle and magnitude of the external alternating magnetic field for the magnetic bubblecade. For this study, the magnetic bubblecade is realized in Pt/Co/Pt films with sizable DMI [15], which have a strong perpendicular magnetic anisotropy (PMA) [18,19]. The bubblecade speed is then examined with respect to the tilt angle and magnitude of the external alternating magnetic field. A clear angular dependence is observed and explained using DW creep theory, which provides an optimal design rule for the magnetic bubblecade.

**Results**

**Schematic Diagram of the Bubble Motion.**

Figure 1 shows a magnetic bubble (up domain) with the Néel DW configuration caused by a positive DMI [15,16,20]. The magnetization (red arrows) inside the DW is pointing radially outward in all directions. By applying a tilted alternating magnetic field, a bubblecade along the +$x$ direction (yellow arrow) was generated [9], of which the measured bubblecade speed $v$ was measured by a magneto-optical Kerr effect (MOKE) microscope with respect to the tilt angle $\theta$ and the amplitude $H$ of the alternating magnetic field. The tilt angle of the electromagnet is defined from the +$z$ direction to the +$x$ direction is shown in Fig. 1.

**Angle Dependence of Bubble Motion.**

Figure 2(a) plots the measured $v$ with respect to $\theta$ under several fixed $H$ as denoted inside the plot. It is clear from the figure that each $v$ curve exhibits a maximum at an angle $\theta_0$ as indicated by the purple arrow. Hereafter, $\theta_0$ will denote the angle for the maximum $v$. The measured $\theta_0$ is plotted with respect to $H$ in Fig. 2(b). The inset of Fig. 2(b) shows that the DW speed $V_{\mathrm{DW}}$ exactly follows the DW creep criticality by showing the linear dependence with respect to $H^{-1/4}$ for the case of $\theta = 0$.

In the creep regime, $v$ can be described by an equation [Supplementary information in Ref. 9] as $v = \beta H_x H_z^{-1/4} \exp[-\alpha_0 H_z^{-1/4}]$, where $\beta$ is a constant related to the asymmetry in the DW motion and $\alpha_0$ is the creep scaling constant for $H_x = 0$. The validity of the present formula was confirmed by experiment for the range of $H_x$ smaller than 50 mT. By replacing the strengths of the magnetic field as $H_z \equiv H \cos\theta$ and $H_x \equiv H \sin\theta$, the equation can be rewritten as a function of $H$ and $\theta$ as given by

$$v(H,\theta) = \beta H^{3/4} \sin\theta \, (\cos\theta)^{-1/4} \exp\left[-\alpha_0 H^{-1/4}(\cos\theta)^{-1/4}\right]. \tag{1}$$

The solid lines in Fig. 2(a) show the best fits with Eq. (1). In this fitting, the experimental value of $\alpha_0$ (=6.7 T$^{1/4}$) is used, which was determined from an independent measurement of the DW creep criticality [18,19,21–24]. Therefore, the fitting was done with a single fitting parameter $\beta$. The good conformity supports the validity of the present equation.

For a given $H$, $\theta_0$ can be obtained from the maximization condition with respect to $\theta$ i.e. $\partial v / \partial \theta|_{\theta=\theta_0} = 0$, as

$$(\cos\theta_0)^{1/4}(4\cot^2\theta_0 + 1) = \alpha_0 H^{-1/4}. \tag{2}$$

Since the variation of $(\cos\theta_0)^{1/4}$ is negligibly small in comparison to $(4\cot^2\theta_0 + 1)$ as shown by Fig. 3(a), it is sufficient to approximate $(4\cot^2\theta_0 + 1) \cong \alpha_0 H^{-1/4}$ for the range of the experimental $\theta_0$ (~30°), leading to

$$\theta_0 = \text{acot}\left[\sqrt{(\alpha_0 H^{-1/4} - 1)/4}\right]. \tag{3}$$

The solid line in Fig. 2(b) shows the numerical evaluation of Eq. (3). Though the experimental data appears scattered in comparison to the small variation of $\theta_0$, the solid line accords well with the experimental data. Please note that Eq. (3) does not contain any fitting parameters, because $\alpha_0$ was determined from an independent measurement.

From the creep equation $V_{\text{DW}} = V_0 \exp\left[-\alpha(H_x)H_z^{-1/4}\right]$, one can find a logarithmic dependence $\alpha(H_x)H_z^{-1/4} = \ln(V_0/V_{\text{DW}})$, where $V_0$ is the characteristic DW speed and $\alpha$ is the creep scaling constant. Due to the logarithmic dependence, $\theta_0$ in Eq. (3) is basically a slowly-varying function of $V_{DW}$. Please note that, even if $V_{DW}$ varies by 10 times, $\theta_0$ changes by only a few degrees, as we discuss later.

**Two-Dimensional Contour Map of Bubble Motion with respect to $\theta$ and $H$.**

To further check the validity of the present theory, we measure the two-dimensional contour map of $v(H,\theta)$, which is plotted with respect to $\theta$ and $H$ as shown by Fig. 3(b). The colour contrast is scaled with the value of $\log(v)$ as the scale bar shown on the right lower end. For this plot, $v$ was experimentally measured for each values of $\theta$ and $H$ over the range of $\theta$ from 10 to 65° with 5° step and the range of $H$ from 16 to 24 mT with 1-mT step. In the map, each colour traces each equi-speed contour. Several equi-speed contours are highlighted by the circular symbols, of which the position $(H,\theta)$ indicates the values of $\theta$ and $H$ of the same speed for each equi-speed contour. The purple solid lines plot the prediction from the present model. It is clear from the figure that the model prediction matches well with the experimental results, confirming the validity of the present theory.

**Dependence of $\theta_0$ on $H$ and $\alpha_0$.**

Figure 4 examines the dependence of $\theta_0$ on $H$ and $\alpha_0$. The circular symbols are obtained by solving Eq. (2) numerically and the solid lines are from Eq. (3). The good conformity between the symbols and lines verifies again the validity of Eq. (3). Figure 4(a) plots $\theta_0$ with respect to $H$ for several fixed $\alpha_0$ over the practical range for Pt/Co/Pt films [13,15,18,19,23,24]. The figure shows that, for all the values of $\alpha_0$, $\theta_0$ increases drastically as $H$ increases up to about 3 mT and then, exhibits a slow variation as $H$ increases further. Figure 4(b) plots $\theta_0$ with respect to $\alpha_0$ for several fixed $H$. It is also seen that $\theta_0$ is greatly reduced for the range of small $\alpha_0$, but slow variation for the range of large $\alpha_0$. The present observations provide a general guideline for the optimal $\theta_0$ to be about 30° for practical experimental conditions.

**Discussion**

We would like to mention that $v$ can be affected by the asymmetries caused by other mechanisms such as chiral damping [16,25,26] or DW width variation [27]. Since the DW width has a dependence on $H$ and $\beta$ is proportional to the DW width, the value of $\beta$ also varies with respect to $H$. However, it is confirmed for the present films that the DW width variation is small (< 30%) [28] and that the chiral damping can be ignored owing to the experimental observation of parabolic $v$ dependence on $H_x$ [15]. The good conformity of the present model to the experimental results reciprocally verifies that the asymmetry of the present films is mainly governed by the DW energy variation and thus, the present films provide a good test system to examine the magnetic bubblecade caused by the DMI-induced asymmetries. Other films with large effects from the other mechanisms [24,25–27,29] require further investigation for each optimal configuration.

Finally, we like to mention that the bubblecade can be realized even in the depinning and flow regimes, where a similar asymmetric DW motion appears [16]. Though we expect that there also exists angular dependence of the maximum bubblecade speed, here we limit our analytic description only in the creep regime, since the origin of the asymmetry in the other regimes is not fully understood yet. However, a similar approach can be applicable to those regimes, once the analytic formula on $H_x$ dependence is uncovered.

In conclusion, we examined the optimal configuration of the external magnetic field for the magnetic bubblecade. From the clear angular dependence of the bubblecade speed, the optimal angle for the maximum speed was determined experimentally and explained theoretically by a model based on the DW creep theory. The optimal angle is finally given by a simple equation of the amplitude of the alternating magnetic field. Our findings directly elucidate the major factors on the dynamics in the magnetic bubblecade, enabling the design of the optimal device configuration.

**Methods**

**Sample preparations.** For this study, Pt/Co/Pt films with strong perpendicular magnetic anisotropy (PMA) were prepared [18]. The detailed layer structure is 5.0-nm Ta/2.5-nm Pt/0.3-nm Co/1.0-nm Pt, which was deposited on a Si wafer with 100-nm $SiO_2$ by use of dc magnetron sputtering. All the films exhibit clear circular domain expansion with weak pinning strength [9,18]. This film has sizeable DMI, which induces asymmetric DW motion.

**Measurement of the bubble speed.** The magnetic domain images were then observed by use of a MOKE microscope equipped with a charge-coupled device camera. To apply a tilted magnetic field onto the films, a Ferris-wheel-like electromagnet is mounted to the microscope, such that it revolves on the *x-z* plane around the focal point of the microscope. The magnetic field can be varied up to 35 mT on the focal plane. The tilt angle of the electromagnet can be controlled from 0 to 90° in 5° steps. To measure the bubblecade speed $v$, a magnetic bubble was initially created by use of the thermomagnetic writing technique [9,14,19]. To apply alternating magnetic field, a magnetic field pulse of $+H$ with a duration time $\Delta t$ is applied with an angle $\theta$ and successively, a reversed magnetic field pulses of $-H$ with the same $\Delta t$ and $\theta$ is applied. After application of each field pulse, the domain image is captured by the MOKE microscope. The bubblecade speed is calculated by measuring the center position of the bubble in each image.

**Figure Captions**

**Figure 1. Schematic descriptions of the unidirectional bubble motion induced by tilted alternating magnetic field.** Illustration of a bubble domain (bright circle) and the DW (grey ring), surrounded by a domain of opposite magnetization (dark area). The red symbols and arrows indicate the direction of the magnetization inside the DW and domains. The dashed circles represent the previous bubble positions and the yellow arrow indicates the direction of the bubble motion.

**Figure 2. Angle dependence of magnetic bubble motion** (a) Measured $v$ with respect to $\theta$ for several $H$ (symbols). The solid lines are best fits with Eq. (1). The purple arrow represents $\theta_0$. The inset plots $V_{\mathrm{DW}}$ with respect to $H^{-1/4}$ for the case of $\theta = 0$. (b) $\theta_0$ with respect to $H$. The solid line is the numerical evaluation of Eq. (3) with the experimental value of $\alpha_0$ (=6.7 T$^{1/4}$).

**Figure 3. Simplification of $(\cos\theta_0)^{1/4}(4\cot^2\theta_0 + 1)$ and two-dimensional contour map of $v(\theta, H)$ as a function of $\theta$ and $H$.** (a) Numerical calculations of $(\cos\theta_0)^{1/4}$, $(4\cot^2\theta_0 + 1)$, and $(\cos\theta_0)^{1/4}(4\cot^2\theta_0 + 1)$ as a function of $\theta_0$. (b) Two-dimensional contour map of $v(\theta, H)$ plotted with respect to $\theta$ and $H$. The colour contrast represents the value of $\log(v)$ with scale bar on the right lower end. The purple solid line is the numerical evaluation of Eq. (3).

**Figure 4. The dependence of $\theta_0$ as a function of $H$ and $\alpha_0$.** (a) $\theta_0$ with respect to $H$ for various $\alpha_0$ and (b) $\theta_0$ with respect to $\alpha_0$ for various $H$. The circular symbols are obtained by solving Eq. (2) numerically and the solid lines are from Eq. (3). The open symbols indicate the experimental data.

**Acknowledgements**

This work was supported by a National Research Foundations of Korea (NRF) grant that was funded by the Ministry of Science, ICT and Future Planning of Korea (MSIP) (2015R1A2A1A05001698 and 2015M3D1A1070465). D.-H.K. was supported by a grant funded by the Korean Magnetics Society. K.W.M. and C.H. were supported from the National Science Foundation Grant No. DMR-1504568, Future Materials Discovery Program through the National Research Foundation of Korea (Grant No. 2015M3D1A1070467). B.-C.M. was supported by the KIST institutional program and Pioneer Research Center Program of MSIP/NRF (2011-0027905).

**Author contributions**

D.-H.K. and K.-W.M. planned and designed the experiment and S.-B.C. supervised the study. D.-H.K. and D.-Y.K. carried out the measurement. S.-C.Y. and B.-C.M. prepared the samples. D.-H.K., K.-W.M., S.-B.C., and C.H. performed the analysis and D.-H.K. and S.-B.C. wrote the manuscript. All authors discussed the results and commented on the manuscript.

**Additional information**

Correspondence and request for materials should be addressed to S.-B.C.

**Competing financial interests**

The authors declare no competing financial interests.

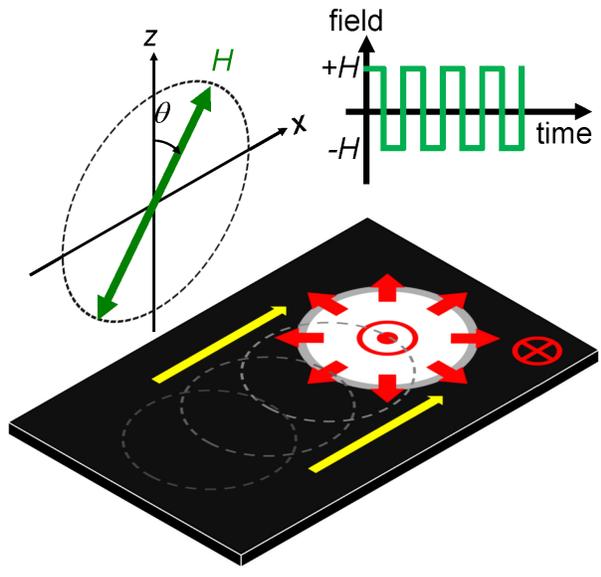

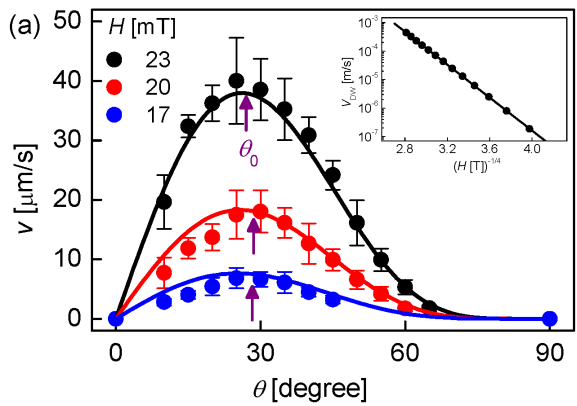

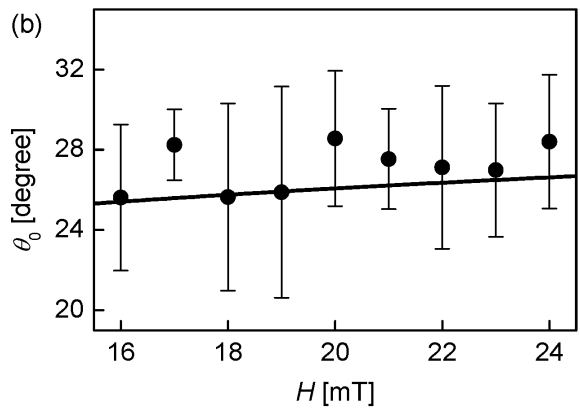

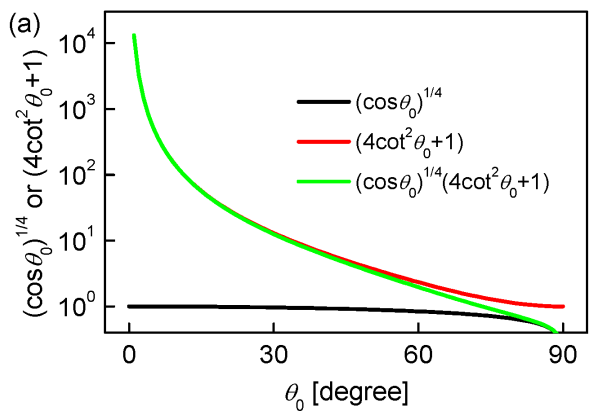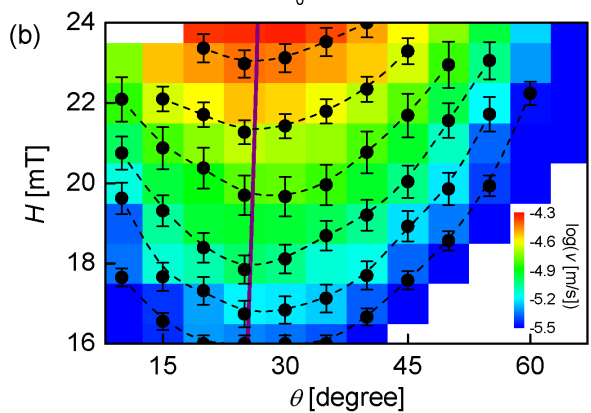

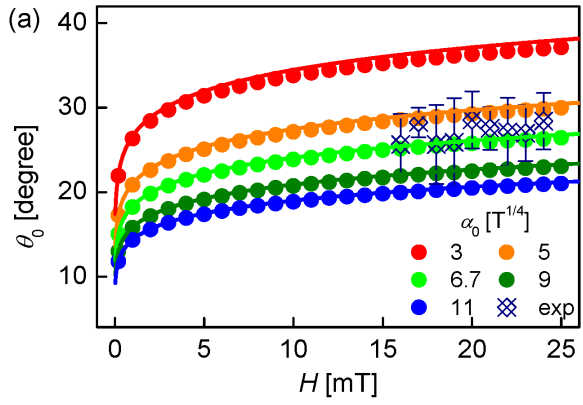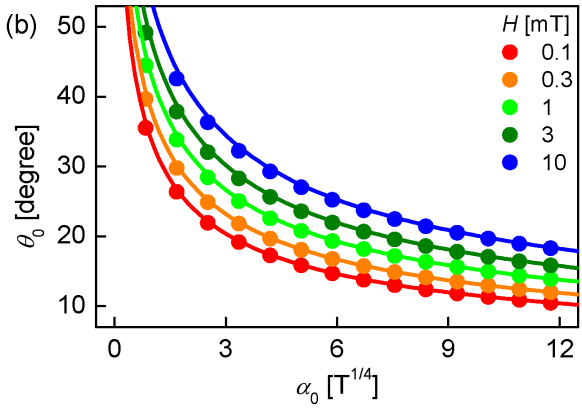